\documentclass[aps,floats]{revtex4}
\usepackage{amsmath}
\usepackage{graphicx,epsfig}

\begin{document}
\bibliographystyle {plain}

\def\oppropto{\mathop{\propto}} 
\def\opsimeq{\mathop{\simeq}}
\def\opoverderline{\mathop{\overline}}
\def\operarrow{\mathop{\longrightarrow}}
\def\opsim{\mathop{\sim}}

\def\fig#1#2{\includegraphics[height=#1]{#2}}
\def\figx#1#2{\includegraphics[width=#1]{#2}}


\title{ Spin-glass chain in a magnetic field :
influence of the disorder distribution  \\
on ground state properties and low-energy excitations  } 

 \author{ C\'ecile Monthus and Thomas Garel}
 \affiliation{Service de Physique Th\'eorique, 
  Unit\'e de recherche associ\'ee au CNRS, \\
  DSM/CEA Saclay, 91191 Gif-sur-Yvette, France}

\begin{abstract}
For the spin-glass chain in an external field $h$, a 
non-zero weight at the origin of the bond distribution $\rho (J)$
is known to induce a non-analytical magnetization 
at zero temperature : for 
$\rho(J) \sim A \vert J \vert^{\mu-1}$ near $J \to 0$,
the magnetization follows the Chen-Ma scaling $M \sim h^{ \mu/(2+\mu)}$.
In this paper, we numerically revisit this model to obtain 
detailed statistical information on the ground state configuration
and on the low-energy two-level excitations
that govern the low temperature properties.
The ground state consists of long unfrustrated intervals
separated by weak frustrated bonds: We accordingly compute the
strength distribution of these frustrated bonds, as well as the
length- and magnetization- distributions of the unfrustrated intervals.
We find that the low-energy excitations are of two types (i) one
frustrated bond of the ground state may have two positions that are
nearly degenerate in energy (ii) two neighboring frustrated bonds of
the ground state may be annihilated or created with nearly zero energy
cost. For each excitation type, we compute its probability density as
a function of its length. We moreover show that the contributions of
these excitations to various observables (specific heat,
Edwards-Anderson order parameter, susceptibility) are in full
agreement with 
direct transfer matrix evaluations at low temperature. 
Finally, following the special bimodal case $\pm J$, where a Ma-Dasgupta
RG procedure has been previously used to compute explicitly
the above observables, we discuss the possibility of an extended
RG procedure : we find that the ground state can be seen as the result
of a hierarchical `fragmentation' procedure that we describe.

\end{abstract}

\maketitle

\section{Introduction}

In this paper, we consider the one dimensional spin-glass chain in a small
external field $h>0$
\begin{eqnarray}
H= - \sum_i J_i \sigma_i \sigma_{i+1} - h \sum_i \sigma_i
\label{model}
 \end{eqnarray}
to obtain detailed results on the ground state and the low-energy
excitations, as a function of the exponent $\mu>0$ characterizing the
weight of the coupling distribution for small couplings
\begin{eqnarray}
\rho(J) \opsimeq_{J \to 0} A \vert J \vert^{\mu-1}
\label{rhoja}
 \end{eqnarray}
As is well-known, the previous model is equivalent to a 
random-bond and random-field ferromagnetic chain \cite{chenma}
\begin{eqnarray}
H= - \sum_i \vert J_i \vert S_i S_{i+1} - h \sum_i x_i S_i
\label{model2}
 \end{eqnarray}
where $x_i=\prod_{j=1}^i {\rm sgn} (J_j) $.

\subsection { Bimodal distribution $J_i=\pm J$: 
Imry-Ma argument and real-space RG}

For the special case of the bimodal distribution $J_i=\pm J$ with probabilities $(1/2,1/2)$, the model (\ref{model2}) corresponds to a pure Ising chain
$\vert J_i \vert=J$ in a bimodal random field $h_i=h x_i=\pm h$.
The Imry-Ma argument \cite{imryma} for the random field Ising chain can be immediately translated for the spin-glass in external field, since the domain walls
of the RFIM now becomes frustrated bonds for the spin-glass :
the random magnetization $m$ of an unfrustrated domain of length $l$
is of order $m \sim \sqrt{l}$, i.e. it gives rise to an energy of order $2 h \sqrt{l}$
in the external field $h$, whereas a pair of two frustrated bonds has
for energy cost $4 J$.
As a consequence, the ground state is made of unfrustrated domains
having the typical Imry-Ma  length $L_{IM} \sim 4 \ J^2/h^2$. 
The real-space Ma-Dasgupta RG \cite{us_rfim} allows to
construct explicitly the positions of frustrated bonds 
and to compute various statistical properties, such as the distribution
of the domain lengths. This approach moreover yields
 the statistics of low-energy two-level excitations \cite{twolevel}.
A natural question is then : what are the corresponding results
for a general distribution $\rho(J)$ that is not bimodal ? 
It turns out that a different behavior occurs if $\rho(J)$
has some weight at small couplings $J \sim 0$. This case, which
includes the Gaussian distribution, completely changes the physics of
the model, as we now discuss. 

\subsection { Distributions with small couplings $\rho(J) \simeq A \vert J \vert^{\mu-1}$ : Chen-Ma argument }

For distributions presenting some weight at small couplings (\ref{rhoja})
the above Imry-Ma argument for the bimodal case is replaced by
the following Chen-Ma argument \cite{chenma}. The essential idea is
that frustrated bonds will be now located on weak bonds, in contrast with
the bimodal case where the cost of a frustrated bond is the same everywhere.
More precisely, the Chen-Ma (CM) argument is as follows : the bonds $J_i$ weaker
than some cut-off $ \vert J_i \vert \leq J_{CM}$ 
are separated by a typical distance of order 
\begin{eqnarray}
l_{CM} \sim J_{CM}^{- \mu}
 \end{eqnarray}
The magnetization of the unfrustrated domain
between two such weak bonds is of order
\begin{eqnarray}
m_{CM} \sim \sqrt{l_{CM}} \sim J_{CM}^{- \mu/2}
 \end{eqnarray}
The flipping of such a domain thus involves a typical energy of order
$J_{CM}$ for the creation of two weak frustrated bonds, but allows to gain a magnetic energy of order $h m_{CM} \sim h J_{CM}^{- \mu/2}$. The balance between the two terms yields an optimal cut-off of order \cite{chenma}
\begin{eqnarray}
J_{CM} \sim h^{ 2/(2+\mu)}
\label{scalingjc}
 \end{eqnarray}
so that the magnetization per spin $M_s$ presents the following non-analytical behavior
\begin{eqnarray}
M_s \sim \frac{m_{CM}}{l_{CM}} \sim h^{ \frac{\mu}{ (2+\mu) }}
\label{scalingms}
 \end{eqnarray}
The zero-temperature susceptibility
\begin{eqnarray}
\chi(T=0) \sim \frac{M_s}{h} \sim h^{-2/(2+\mu)}
 \end{eqnarray}
thus diverges at zero field $h \to 0$.
For instance, the Gaussian distribution $\rho(J)$,
which corresponds to the exponent $\mu=1$, leads to the behavior $M_s \sim h^{1/3}$. 
The critical exponent $\frac{\mu}{ (2+\mu) }$ for the magnetization in 
external field
 was found to be exact via transfer matrix calculations
by Gardner and Derrida \cite{gardner}, 
where the prefactor was moreover computed.
The presence of small couplings does induce interesting new properties
for the ground state with respect to the bimodal case.

Chen and Ma have also analyzed the low temperature properties,
in particular in the regime where the temperature $T$ is much smaller than the typical energy $J_{CM}$ of a domain.
In this regime $T \ll J_{CM}$ (\ref{scalingjc}), only a small fraction of two-level excitations
will be excited : the density $\rho(E=0)$ 
of excitations near zero energy can be estimated to scale as \cite{chenma}
\begin{eqnarray}
\rho(E=0) \sim \frac{1}{l_{CM}} 
\times \frac{1}{J_{CM}} \sim J_{CM}^{\mu-1} \sim 
h^{ 2(\mu-1)/(2+\mu)}
\label{rhozero}
 \end{eqnarray}
where $1/l_{CM}$ represents the density of frustrated bonds in the ground state,
and where $1/J_{CM}$ represents the fraction of these frustrated bonds
 that will be involved in excitations of vanishing energy $E \to 0$. This excitations density is then expected to govern the
leading term of the specific heat at low temperature
 \cite{anderson72}
\begin{eqnarray}
C \opsimeq_{T \to 0} T \frac{\pi^2}{6} \rho(E=0) +O(T^2)
\label{crho}
 \end{eqnarray}

In this paper, our aim is to study in some details the statistical properties
of the ground state configuration, and
of the the low-energy excitations that govern the low-temperature properties.

\section{ Statistical properties of the ground state configuration }

In this section, we present the numerical results that we have obtained
via the zero-temperature transfer-matrix formulation \cite{gardner}
from which one can obtain the ground state configuration
$\{\sigma_i \}$ in each given sample $\{J_i \}$.
In the whole paper, we have used the following distribution for the couplings
\begin{eqnarray}
\rho( J )=\frac{\mu}{2} \vert J \vert^{\mu-1} \ \ \ {\rm for} \ \ -1
\leq J \leq 1 
\label{rhoj}
 \end{eqnarray}
yielding $A=\frac{\mu}{2}$. The results given in this section have
been obtained from averages over $10^5$ independent chains containing
$N \sim 10^6$ sites.

\subsection{Magnetization per spin }

\begin{figure}
\centerline{\includegraphics[height=8cm]{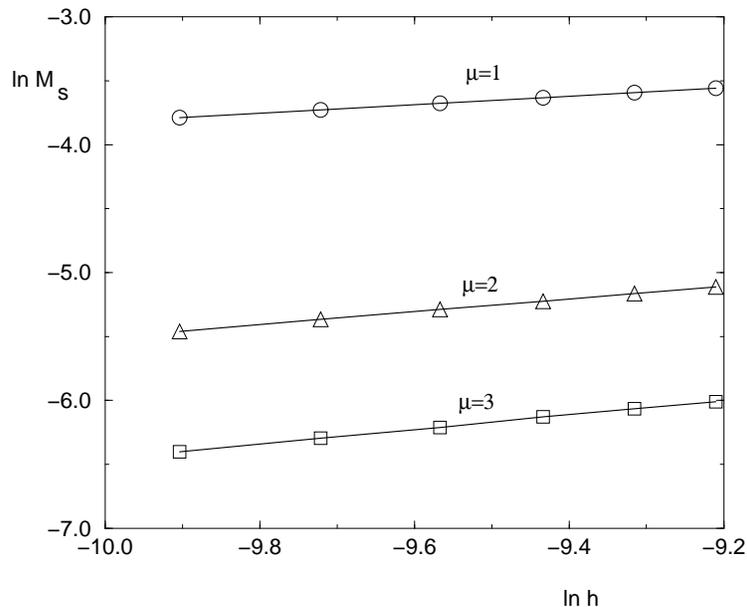}} 
\caption{ Magnetization per spin in the ground state as a function of the external field varying between $h=5. 10^{-5}$ and $h=1. 10^{-4}$ in log-log plot, for  disorder distributions (\ref{rhoj})
with exponents $\mu=1, 2 , 3$.
The corresponding slopes are in full agreement with the 
exact exponents $1/3$, $1/2$, $3/5$ ( Eq \ref{resexact}).  } 
\label{scalingm}
\end{figure}

As a first observable, we have computed the magnetization per spin
which corresponds to a thermodynamic quantity which is exactly
known from a transfer matrix calculation done by Gardner and Derrida who have obtained \cite{gardner}
\begin{eqnarray}
M_s^{exact} \opsimeq_{h \to 0} (\mu+1) \left( \frac{ 4  A}{ \mu (\mu+2)^2} \right)^{\frac{1}{\mu+2}} \frac{ \Gamma \left( \frac{\mu+1}{\mu+2} \right)}{ \Gamma \left( \frac{1}{\mu+2} \right)}
  h^{ \frac{\mu}{ (2+\mu) }}
\label{resexact}
 \end{eqnarray}
 We have checked that both the scaling in $h$ (see Figure \ref{scalingm})
and the prefactor are in excellent
agreement with the exact result (\ref{resexact}). 

Now that we have identified the regime in $h$ where the scaling
(\ref{scalingms}) is satisfied, we may turn to more refined statistical properties for which, to the best of our knowledge, no exact expression is available.

\subsection{Probability distribution of frustrated links }

\begin{figure}
\centerline{\includegraphics[height=8cm]{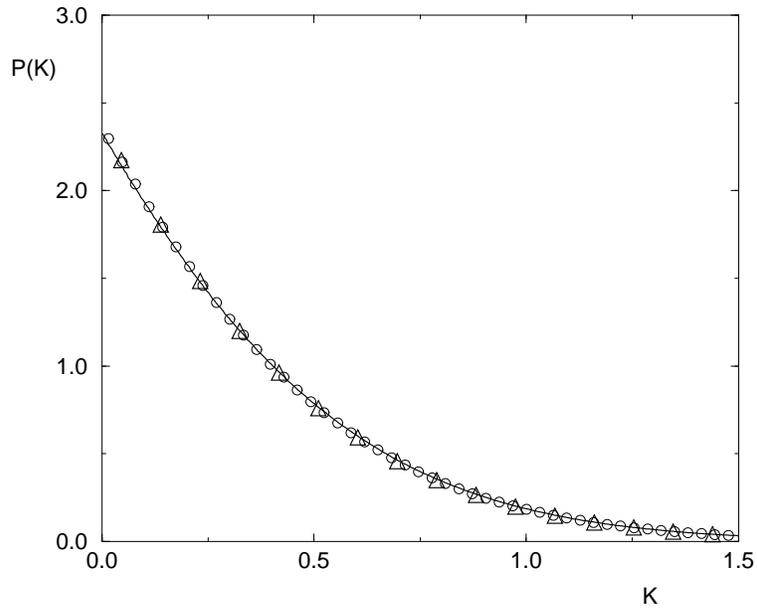}} 
\caption{ 
Rescaled probability distributions $P_{\mu=1}(K=\vert J \vert/J_{CM}(h))$
of frustrated links in the ground state : the data for various fields,
namely $h=1. 10^{-2}$ (line),
 $h=5. 10^{-4}$ (circles) and  $h=1. 10^{-4}$ (triangles),
follow the same master curve.  } 
\label{scalingjfrustrated}
\end{figure}

We have computed the normalized probability density of coupling $\vert J \vert$
among frustrated links
\begin{eqnarray}
P^f(J)= \frac{ N_f(J) }
{ \int dJ' N_f(J')  }
 \end{eqnarray}
where $N_f(J)$ represents the number of frustrated links of strength $J$.

According to the Chen-Ma argument, the frustrated links should 
have a typical strength of order $J_{CM} \sim h^{2/(2+\mu)}$.
We have thus plotted on Figure \ref{scalingjfrustrated}
 the probability distribution of frustrated links in terms
the appropriate rescaled variable 
\begin{eqnarray}
K = \frac{ \vert J \vert}{ J_{CM}(h) }= \frac{ \vert J \vert}{ h^{2/(2+\mu)}}
\label{scalingJ}
 \end{eqnarray}
for various $h$ with the same initial distribution (\ref{rhoj})
corresponding to $\mu=1$.

\begin{figure}
\centerline{\includegraphics[height=8cm]{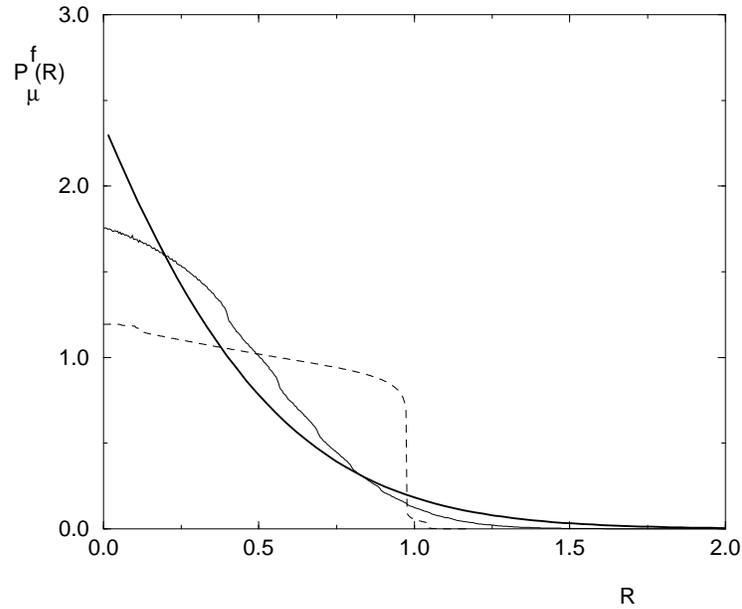}} 
\caption{ 
Probability distributions $P_{\mu}^f(R \equiv r/r_{CM}(h) = 
(\vert J \vert/J_{CM}(h))^{\mu}$)
of the frustrated links for various disorder distributions (\ref{rhoj})
corresponding to the exponents $\mu=1$ (bold line), $\mu=0.5$ (thin line) and $\mu=0.1$ (dashed line). } 
\label{scalingrfrustrated}
\end{figure}

To compare the distributions of frustrated links corresponding
to different disorder distributions $\rho(J)$ characterized by 
different exponents $\mu$ (\ref{rhoj}), it is more convenient
to consider the reduced variable 
\begin{eqnarray}
r=\vert J \vert^{\mu}
 \end{eqnarray}
which is distributed with the flat distribution
\begin{eqnarray}
\pi^{a priori} (r) = \theta(0 \leq r \leq 1)
 \end{eqnarray}
for any $\mu$ in (\ref{rhoj}).
Taking into account the Chen-Ma scaling
 \begin{eqnarray}
r_{CM}(h)=J_{CM}^{\mu}(h) =h^{2 \mu /(2+\mu)}
 \end{eqnarray}
we have plotted on Figure \ref{scalingrfrustrated} 
the probability distribution $P_{\mu}^f(R \equiv r/r_{CM}(h))$
of the frustrated links for various disorder distributions
corresponding to the exponents $\mu=1$, $\mu=0.5$ and $\mu=0.1$ :
for $\mu=1$, this distribution is rather smooth, whereas it becomes 
steeper as $\mu$ decays. In particular, for $\mu=0.1$, it 
becomes close to a simple theta function $\theta (R \leq 1)$.
This shows that in the limit $\mu \to 0$, the properties of the ground state become simpler, and we will discuss this point in more details in Appendix \ref{mutozero}.

\subsection{Probability distribution of the lengths of unfrustrated intervals }

\begin{figure}
\centerline{\includegraphics[height=8cm]{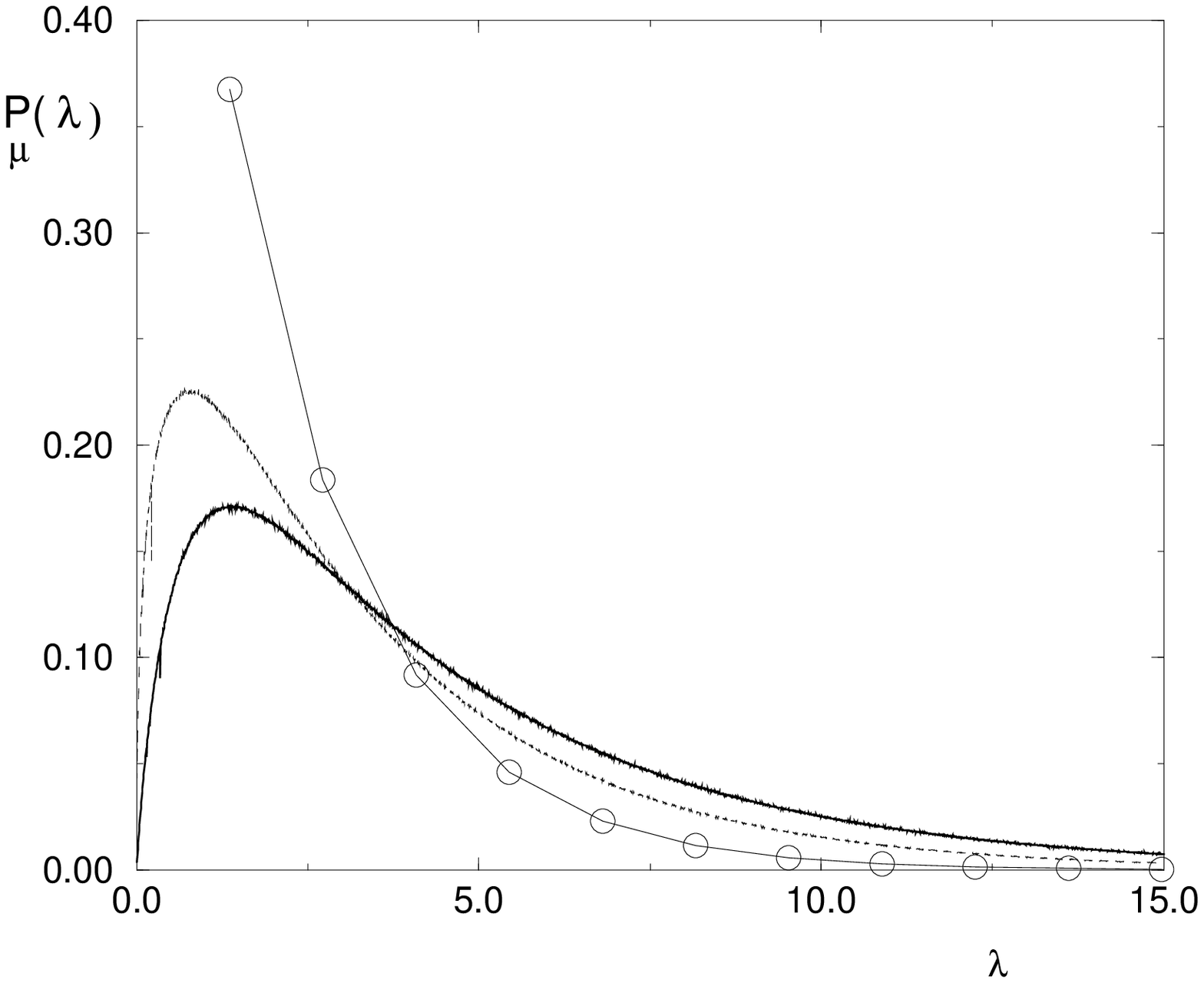}} 
\caption{ 
Rescaled probability distributions ${\cal P}_{\mu}(\lambda= l /l_{CM(h)})$
of the length between two frustrated bonds of the ground state, for various initial disorder distribution characterized by the exponent  $\mu=1$
(full line) $\mu= 1/2$ (dashed line), $\mu=1/4$ (circles)  } 
\label{scalinglunfrustrated}
\end{figure}

According to the Chen-Ma argument, the length $l$ between two
frustrated links has for typical scale $l_{CM} \sim h^{-2 \mu /(2+\mu)}$.
Indeed, we obtain that the appropriate rescaled variable for the length
of unfrustrated intervals is
\begin{eqnarray}
\lambda =\frac{ l}{ l_{CM}(h) }= l h^{2\mu /(2+\mu)}
 \end{eqnarray}
as h varies.
The probability distribution ${\cal P}_{\mu}(\lambda)$ 
of the scaling variable $\lambda$ is plotted on Figure \ref{scalinglunfrustrated}
for various $\mu$.

For $\lambda \to 0$, in contrast with the bimodal case where
${\cal P} (\lambda=l/l_{IM}) $ presents an essential singularity \cite{us_rfim}, 
we obtain here power-law behavior
near the origin
\begin{eqnarray}
{\cal P}_{\mu}(\lambda) \oppropto_{\lambda \to 0} \lambda^{\alpha(\mu)}
 \end{eqnarray}
with an exponent $\alpha(\mu)$ that grows with $\mu$ (see Figure \ref{scalinglunfrustrated}).
For instance for $\mu=1$, the best fit yields the exponent $\alpha(\mu=1) \simeq 0.8$.
For large $\lambda$, the decay is exponential
\begin{eqnarray}
{\cal P}_{\mu}(\lambda) \oppropto_{\lambda \to \infty} e^{-\gamma(\mu) \lambda }
 \end{eqnarray}
For $\mu=1$, the best fit yields $\gamma(\mu=1) \simeq 0.25$.

\subsection{Probability distribution of the magnetizations of unfrustrated intervals }

\begin{figure}
\centerline{\includegraphics[height=8cm]{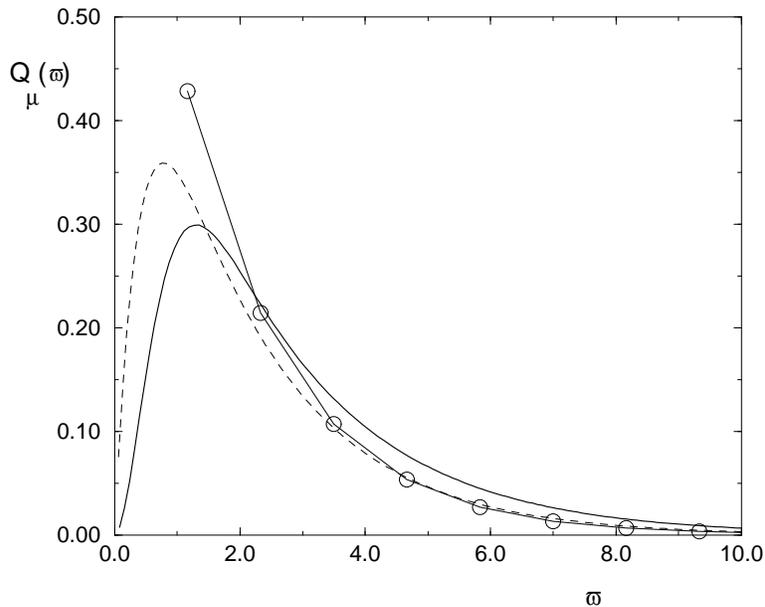}} 
\caption{ 
Rescaled probability distributions $Q_{ \mu}(\omega= m /m_{CM}(h))$
of the magnetization between two frustrated bonds of the ground state, for various initial disorder distribution characterized by the exponent  $\mu=1$
(full line) $\mu= 1/2$ (dashed line), $\mu=1/4$ (circles). } 
\label{scalingmunfrustrated}
\end{figure}

Similarly, we find that, in agreement with
the Chen-Ma argument, the appropriate rescaled variable for the magnetization
of unfrustrated intervals is
\begin{eqnarray}
\omega = \frac{ m}{ m_{CM}(h) }=m h^{\mu /(2+\mu)}
 \end{eqnarray}
as $h$ varies.
The probability distribution $Q_{\mu}(\omega)$
of the scaling variable $\omega$ is plotted  
on Figure \ref{scalingmunfrustrated}
for various $\mu$.

Again, for $\omega \to 0$, we obtain power-law behaviors
\begin{eqnarray}
Q_{\mu}(\omega) \oppropto_{\omega \to 0} \omega^{\beta(\mu)}
 \end{eqnarray}
with for instance $\beta(\mu=1) \sim 1.5$, whereas the decay for large $\omega$ is exponential
\begin{eqnarray}
Q_{\mu}(\omega) \oppropto_{\omega \to \infty} e^{-\delta(\mu) \omega }
 \end{eqnarray}
with $\delta(\mu=1) \simeq 0.5$.

\section{Low-temperature properties }

\subsection{ Low temperature transfer-matrix results for various observables }

\begin{figure}
\centerline{\includegraphics[height=8cm]{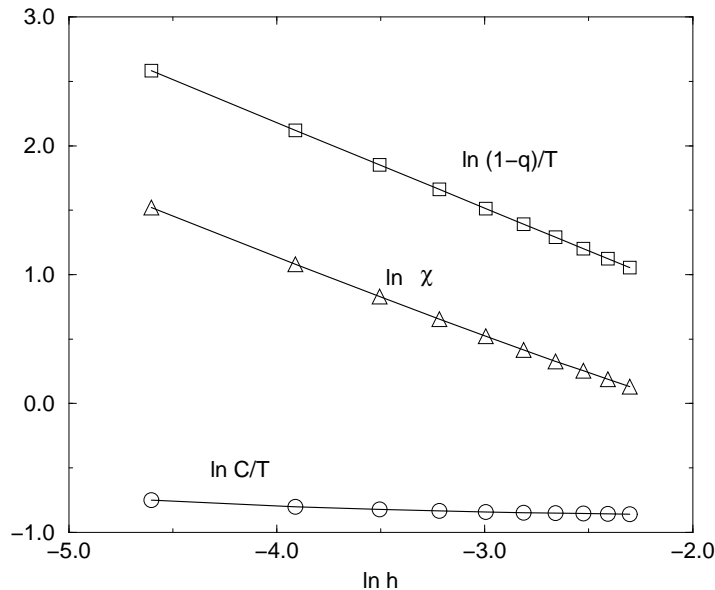}} 
\caption{ 
Low temperature behaviors of the specific heat $C$, of the Edwards-Anderson
order parameter $q$ and of the susceptibility $\chi$ as a function of the
external field in log-log plot for the special value $\mu=1$ :
the exponents are respectively $-0.04$ for the specific heat,
$-0.66$ for $q$ and $-0.6$ for $\chi$ (see text for more details).  } 
\label{chaleuretcie}
\end{figure}

We have first computed via transfer matrix
various observables in the low-temperature regime $T \ll J_{CM}(h)$
(\ref{scalingjc}). For instance for $h=0.02$ corresponding to
$J_{CM}(h)=0.07$, we have checked, 
for temperatures $T=2. 10^{-3}$, $T=3. 10^{-3}$, $T=4. 10^{-3}$, $T=5. 10^{-3}$,
that the leading term of the specific heat is linear in $T$
\begin{eqnarray}
C \equiv \frac{ < E_N^2> -<E_N>^2}{ T^2 N} \oppropto_{T \to 0} T 
\label{chaleurt}
 \end{eqnarray}
and that the leading term of the susceptibility is a constant
\begin{eqnarray}
\chi(T) \equiv \frac{ < M_N^2> -<M_N>^2}{ T N} \oppropto_{T \to 0} cte
\label{suscept}
 \end{eqnarray}
We have also found that the Edwards-Anderson order parameter
deviates linearly in temperature from the zero-temperature value
$q_{EA}(T=0)=1$
\begin{eqnarray}
q_{EA} \equiv \overline{<\sigma_i>^2 } \opsimeq_{T \to 0} 1- T (cte') 
 \end{eqnarray}

We have then studied the dependence in the external field $h$ at fixed temperature.
For instance, for $T=5.10^{-3}$, we have studied the dependence
in $h=10^{-2}$ to $h=10^{-1}$ (see Figure \ref{chaleuretcie}).
The results for the specific heat is in good agreement with the
Chen-Ma prediction 
(Eq \ref{crho} and \ref{rhozero})
\begin{eqnarray}
\frac{C}{T} \opsimeq_{T \to 0} h^{2(\mu-1)/(2+\mu)}
\label{chaleur}
 \end{eqnarray}
as well as the susceptibility
\begin{eqnarray}
\chi(T) \opsimeq_{T \to 0} h^{-2/(2+\mu)}
\label{suscep}
 \end{eqnarray}

We have also computed the Edwards-Anderson order parameter, and
found the same exponent as for the susceptibility (\ref{suscep})
\begin{eqnarray}
\frac{1-q_{EA}}{T} \opsimeq_{T \to 0} h^{ -2/(2+\mu)} 
 \end{eqnarray}
which can be explained from the analysis in terms of the 
low-energy two-level excitations, as we now explain.

\subsection{ Interpretation in terms of low-energy two-level excitations }

We have already given the expression (\ref{crho}) of the specific heat
in terms of the density $\rho(E=0)$ of excitations near zero energy.
The similar expressions for the Edwards-Anderson order parameter
and the susceptibility read \cite{twolevel}
\begin{eqnarray}
\frac{1-q_{EA}}{T}\opsimeq_{T \to 0} 2 \int_0^{+\infty} dl \ l \rho(E=0,l)
\label{qrho}
 \end{eqnarray}
\begin{eqnarray}
\chi\opsimeq_{T \to 0} 2 \int_{-\infty}^{+\infty} dm \ m^2 \rho(E=0,m)
\label{chirho}
 \end{eqnarray}
where $\rho(E=0,l)$ represents the density of excitations of length $l$,
and where $\rho(E=0,m)$ represents
 the density of excitations of magnetization $m$.
Note that in the random field chain \cite{twolevel}, the magnetization
of a ferromagnetic domain is equal to its length, whereas 
here in the spin-glass chain, it is not the case, since the magnetization of an unfrustrated domain
scales as $\sqrt{l}$. This is why the scaling in $h$ are the same
here for these two observables.

\begin{figure}
\centerline{\includegraphics[height=8cm]{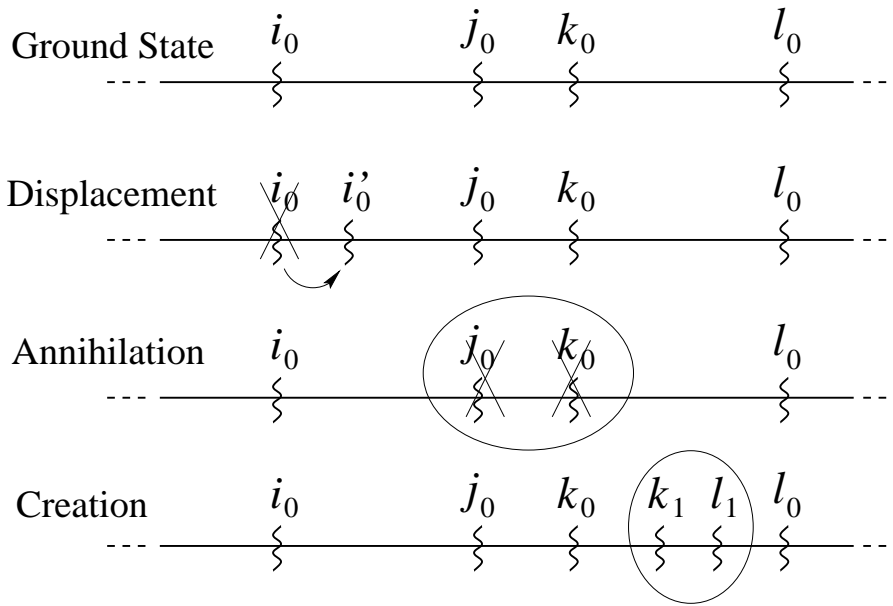}} 
\caption{ 
Nature of the low-energy excitations. The ground state is made of long  
unfrustrated intervals separated by frustrated bonds called
$.. i_0,j_0,k_0,l_0 ...$. Low energy excitations may be of several types
(displacement, annihilation or creation of frustrated bonds). } 
\label{defexcita}
\end{figure}

We have numerically computed the probability density of excitations,
as a function of their size $l$ and of their type.
Indeed, we obtain by an exhaustive numerical enumeration
that the total density of low-energy
excitations is exactly the sum of three contributions

\begin{eqnarray}
\rho_{tot}(E=0,l)= \rho_{disp}^{(1)}(E=0,l) + \rho_{anni}^{(2)}(E=0,l)+ \rho_{crea}^{(2)}(E=0,l)
\label{decomposition}
 \end{eqnarray}

(1) The excitations of type 1, of density $\rho_{disp}^{(1)}(E=0,l)$,
 involve a single frustrated link of the ground state,
of coupling $J_a$, 
that can be displaced to another position of coupling $J_b$
(inside the intervals defined by the two frustrated neighbors of $J_a$)
 with almost no energy cost. The energy difference
\begin{eqnarray}
\Delta E^{(1)}= - 2 \vert J_{a} \vert + 2 \vert J_{b} \vert
 + 2 h m_{ab}      \leq T
 \end{eqnarray}
involves the defrustration of $J_a$, the frustration of $J_b$ and
the magnetization $m_{ab}$ between these two links.

\begin{figure}
\centerline{\includegraphics[height=8cm]{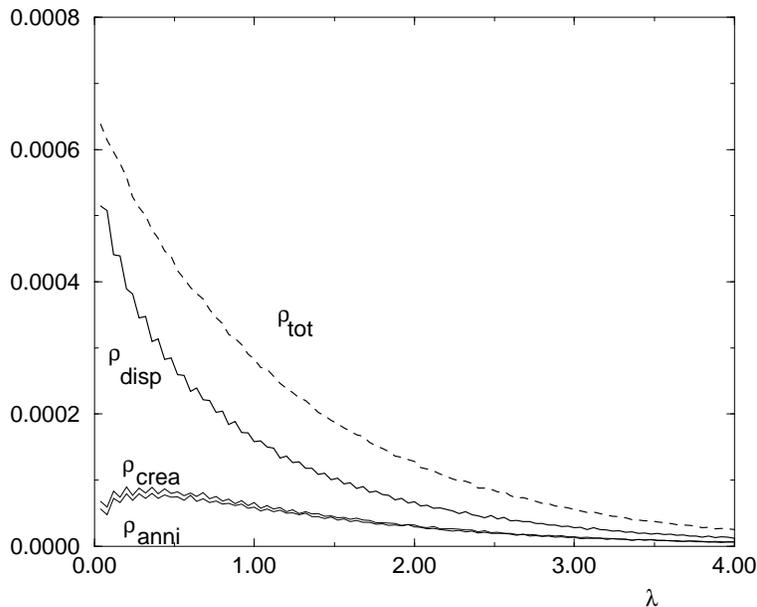}} 
\caption{ 
Densities of the three types of low energy excitations present in
equation (\ref{decomposition}), in terms of 
their rescaled length $\lambda=l/l_{CM}(h)$ in the case $\mu=1$.  } 
\label{rhoexcita}
\end{figure}

(2) The excitations of type 2 involves a pair of neighbor frustrated bonds
that can be annihilated ($\rho_{anni}^{(2)}(E=0,l)$)
or that can be created ($\rho_{crea}^{(2)}(E=0,l)$) with almost no energy cost
\begin{eqnarray}
\Delta E_{anni}^{(2)} &&=  -2 \vert J_{1} \vert - 2 \vert J_{2} \vert
 + 2 h m_{12}      \leq T \\
\Delta E_{crea}^{(2)} && =  2 \vert J_{1} \vert + 2 \vert J_{2} \vert
 + 2 h m_{12}      \leq T
 \end{eqnarray}
These two type of excitations are symmetric and are thus expected to correspond
to the same distribution
\begin{eqnarray}
\rho_{anni}^{(2)}(E=0,l)= \rho_{crea}^{(2)}(E=0,l)
 \end{eqnarray}
which we have checked in our numerical results.
The densities of these excitations in terms of their
rescaled length $\lambda=l/l_{CM}(h)$ 
are given on Figure \ref{rhoexcita} for $\mu=1$.

We have moreover checked the relations (\ref{crho},\ref{qrho},\ref{chirho})
between, on the one hand, the specific heat, the Edwards-Anderson order
parameter, and the susceptibility obtained from low temperature transfer-matrix
calculations, and on the other hand the total number of excitations,
their averaged length, and their averaged square magnetization.
The agreement shows that the excitations described above are the only ones
that play a role in the low temperature behavior of these observables.
The present analysis in terms of the statistics of low-energy excitations
 thus gives a microscopic interpretation of the low temperature
equilibrium properties.

\section{ Renormalization procedures in each sample}

As already mentioned in the introduction,
in the bimodal case $\pm J$, there exists a real-space RG procedure
that allows to compute exactly the statistical properties of the ground state
\cite{us_rfim} (such as the domain length distribution)
as well as the statistics of low-energy excitations \cite{twolevel}.
A natural question is thus : 
is there a generalized RG procedure that would be valid for the spin-glass
beyond the bimodal case $J_i=\pm J$? 
Before trying to answer this question for the $\rho(J) \sim A \vert J \vert^{\mu-1}$ case,
let us first briefly recall the principle of the RG procedure
for the bimodal case.

\subsection{ Bimodal distribution }

The RG procedure defined in ref \cite{us_rfim} for the bimodal case,
consists in an optimization from small scales towards large scales :
one starts for instance from the completely magnetized state $\sigma_i=+1$,
that contains many frustrated bonds, and one flips iteratively
the unfrustrated domain presenting the smallest magnetization, as long
as the energy is lowered, i. e. as long as the balance 
between the energy gained from the suppression of the two boundary
frustrated bonds is bigger than the energy loss $(2 h m)$ from the negative orientation
with respect to the external field
\begin{eqnarray}
\Delta E^{flip}= -4 J + 2 h \vert m \vert <0
 \end{eqnarray}
So the RG procedure has to be stopped when all unfrustrated domains have
magnetizations $m \geq 2 J/h$ : the state obtained then corresponds to the ground state. 
What makes the renormalization tractable in this case is
that due to the constant cost $(2 J)$ of any domain wall,
the renormalization concerns a one-dimensional potential,
namely the magnetization as a function of the running point.

\subsection{ General distribution : hierarchical RG based on the energy }

In this case, the problem cannot be reformulated as the renormalization
of a one-dimensional potential, since in addition to the magnetization,
one has to take into account that the couplings $J_i$ vary along the chain.
Moreover, we have seen with the Chen-Ma argument that 
the frustrated links are concentrated on small couplings.

As a first step, we can thus formulate the following renormalization
that optimizes from the biggest scales towards smaller scales :

(i) one starts from the state with no frustrated links
that presents a positive magnetization 
(of order $\sqrt{N}$ for a chain of size $N$) : it corresponds to
one of the two mirror zero-field ground states.

(ii) at each step, we flip the interval  $(\sigma_{i+1}, ... \sigma_j)$ 
that allows the maximal decrease of energy $E= \displaystyle \min_{i<j} \Delta E(i,j) <0$
where
\begin{eqnarray}
 \Delta E(i,j) = 2 f(J_i) \vert J_i \vert + 2 f(J_j) \vert J_j \vert
+ 2 h \sum_{k=i+1}^j \sigma_k
 \end{eqnarray}
where $f(J_i)=1$ if $J_i$ becomes frustrated during the flip, $f(J_i)=-1$ if $J_i$ becomes unfrustrated during the flip.
If we find no interval to flip, we have obtained the ground state,
since by definition the ground state is stable upon the flip of any interval.

We have indeed checked that this procedure allows to obtain the exact ground state computed
independently via the zero temperature transfer matrix.
We have moreover obtained that actually, the links that become 
frustrated during the procedure never get `unfrustrated' later
in this procedure. 
And in fact the following hierarchical RG procedure,
where at generation $n$, the chain is cut into a certain number of sub-chains,
gives the exact ground state :

(i) same initial state as before

(ii ) once the first interval $(i_1,j_1)$ to be flipped is found,
we can find the next intervals to be flipped independently
within the three sub-chains $(1,i_1)$, $(i_1,j_1)$ and $(j_1,N)$.
And we iterate until there is no interval to flip anymore.

The fact that the first links $(i_1,j_1)$ that become frustrated
indeed belongs to the final ground state, can be justified 
via a `reductio ad absurdum' \cite{jerome}, and then it is valid
for all stages of the procedure.

This hierarchical procedure thus defines an energy driven fragmentation process
of the chain, whose statistical properties can be studied 
and compared to other fragmentation models \cite{fragmentation}.
In particular, we have studied the number $n_{split}$ of splitting 
and the number $n_{gene}$ of generations as a function of the size $N$
of the chain, for $N=500 $ to $N=8000$.
For the disorder distribution (\ref{rhoj}) with $\mu=1$
and external field $h=0.02$, that corresponds to the length Chen-Ma scale
$l_{CM}(h) \sim 13.5..$, we obtain, as expected, that the number of splittings
grows linearly in $N$
\begin{eqnarray}
\frac{ n_{split}(N) }{ N } \opsimeq_{N \to \infty}  8 \ 10^{-3} 
 \end{eqnarray}
and that the number of generations grows logarithmically in $N$
\begin{eqnarray}
\frac{ n_{gene}(N) }{ \ln N } \opsimeq_{N \to \infty}  2.1
 \end{eqnarray}

This RG analysis reveals a hierarchical structure
among the frustrated links of the ground state.
This hierarchy has both a spatial meaning, but also an energy meaning.
Indeed, since an interval created inside another interval
has, by definition of the RG procedure, a smaller energy,
it is clear that the low-energy excitations of the type `annihilation'
 introduced in (\ref{decomposition}) can only concern a pair
of frustrated bonds that have been created together 
and that have no descendent in the hierarchy.
Similarly, the low-energy excitations of the type `creation'
introduced in (\ref{decomposition}) correspond to a pair 
of frustrated bonds that would have been created next if the procedure
had been applied a bit beyond $\Delta E \leq 0$. 
Finally, the only remaining excitations are the `displacements'
(\ref{decomposition}), that actually also preserve
the hierarchical structure, since a link can move only between
its two frustrated neighbors.

\subsection{ RG based on the weakest link at each step }

Since the frustrated links concentrate
on the links that are weak, i.e. of order $J_{CM}(h)$,
 it is tempting to try to define a RG procedure
based on the weakest link at each step.
In Appendix \ref{mutozero}, we show that a simple RG procedure 
based on this idea becomes exact in the limit $\mu \to 0$,
that corresponds to an infinitely broad distribution (\ref{rhoj})
near $J \to 0$.

\section{ Conclusion }

For the spin-glass chain in an external field $h$, 
 we have first studied via zero-temperature transfer-matrix calculations
the statistical properties of the ground state configuration.
We have then studied the nature and the statistics
of the low-energy two-level excitations, via a direct enumeration,
and we have analyzed their contribution to
the specific heat, the Edwards-Anderson order parameter and the
susceptibility in the low-temperature regime. 
Finally, we have shown that an extended
RG procedure, based on the iterative flipping of the best energetic interval,
could be used to obtain the exact ground state in external field.
This RG procedure reveals a hierarchical structure
among the frustrated links present in the ground state.

The possible relation of this hierarchical picture with higher
dimensional disordered models is clearly of interest.
In the 2d random field Ising model, a spatial hierarchical picture has
been identified long ago \cite{binderrfim}, based on the existence of the Imry-Ma
domain length scale. We tend to think that this hierarchy is energetic
in nature. More precisely, we believe that a RG procedure that would
start from the ferromagnetic pure state, and flip iteratively, at each
step, the most advantageous domain- regardless of its size- will
ultimately converge towards the ground state. After the initial
flipping of the most advantageous domain (which, through 
the Imry-Ma argument, is also the largest), one has to search
separately for the next advantageous domains inside and outside the
initial one. These flipped domains will then display a disjoint or
hierarchically nested structure. Another related problem, where an
iterative optimization procedure starts from the largest scale and
hierarchically proceeds towards smaller scales, has
been studied by Binder \cite{binderinterface} for interfaces.

For the 2d or 3d spin glass case at zero magnetic field, there is no
equivalent of the Imry-Ma or Chen-Ma length scale. The existence of a
hierarchical organization in ground state or low temperature
properties has nevertheless been found along various lines: rigidity
properties \cite{rammal}, distance between spin configurations subject to
the same thermal noise \cite{derrida,arcangelis}, calculations on small 
systems \cite{kobe}, extensive data clustering analysis
\cite{domany}. This hierarchical organization pertains to spin
clusters, and is not a priori linked to a hierarchy 
in their flipping energies \cite{Do_He}.

\section{Acknowledgments }

It is a pleasure to thank B. Derrida and J. Houdayer
for useful discussions, as well as E. Domany and G. Hed for correspondence.

\appendix

\section{ Simple soluble RG procedure in the limit $\mu \to 0$ }

\label{mutozero}

In the limit $\mu \to 0$, the disorder distribution $\rho(J)$ (\ref{rhoj}) becomes 
infinitely broad distribution near $J \to 0$.
 It is thus tempting to define a RG procedure
based on the weakest link at each step.
In the following, we consider the simplified RG procedure :

(i)  same initial state as before : one starts from the state with no frustrated links that presents a positive magnetization 

(ii) First iteration :   we choose the smallest 
coupling in absolute value
  $\Gamma= \vert J_{min} \vert$. The chain is thus decomposed into two
sub-chains. We consider the magnetizations of the two sub-chains
$m_1+m_2=M>0$ : if one of the two magnetizations $(m_1,m_2)$ is
negative, for instance, $m_1<0$, we will flip the sub-chain 1 if it
lowers the energy, i.e. if the balance between the cost $2 \Gamma$ of
introducing a frustrated bond between the two sub-chains is less than
the energy gained by the orientation of the sub-chain 1 along the
external field 
\begin{eqnarray}
\Delta E^{flip}= -2 h \vert m_1 \vert +2 \Gamma <0 
 \end{eqnarray}
otherwise, if $\Delta E^{flip}>0$, we do not flip the sub-chain 1.
After this, the two sub-chains will evolve 
as two independent sub-chains with free boundary conditions,
so we iterate the procedure.

This very simple RG procedure is of course not exact, since at each step,
we neglect the weak bond at the other boundary of the interval
that has been previously decimated $\vert J \vert < \Gamma$.
Indeed, at each step, we consider that the cost of the flipping of an interval is exactly
$2 \Gamma$, whereas it should be $2 (\Gamma \pm \epsilon)$, where 
$\epsilon <\Gamma$ is the absolute value of the coupling at the other boundary,
that has been previously decimated, and where the sign $(\pm)$
depends on the state of this coupling $\epsilon$, frustrated or not, in the
renormalized chain at the RG scale $\Gamma$.
However, we will show below that it becomes exact in the limit
$\mu \to 0$, where the distribution becomes 
infinitely broad distribution near $J \to 0$.

\subsection{Statistical properties of the intervals between weak bonds at RG scale $\Gamma$ }

In this section, we study the RG procedure defined above
in the thermodynamic limit of an infinite chain.  
At the renormalization scale $\Gamma$, the chain is split into
independent unfrustrated intervals, separated by weak bonds that can be either
frustrated or not.
We now derive some statistical properties of these intervals
between weak bonds.

The distribution $P_{\Gamma}(l)$ of the length $(j-i)$ between two weak bonds $J_{i,j}<\Gamma$
is simply exponential, since it corresponds to the probability
that $l$ independent couplings have $\vert J_k \vert \geq \Gamma$,
\begin{eqnarray}
P_{\Gamma} (l) \opsimeq \frac{1}{l_{\Gamma} } e^{- \frac{l}{l_{\Gamma} }}
\label{distril}
 \end{eqnarray}
with the characteristic length
\begin{eqnarray}
l_{\Gamma} = \frac{1}{- \ln \left[ 1-\int_{\vert J \vert \leq \Gamma} dJ \rho(J) \right]} \opsimeq_{\Gamma \to 0}  \frac{\mu}{ 2 A \Gamma^{\mu} }
\label{lg}
 \end{eqnarray}

The magnetization of an unfrustrated interval of length $l$ is simply the sum
$m= \pm \sum_{i=1}^l sgn(J_i)$.
Since $l$ is large, the distribution of $m$ given $l$ is 
a Gaussian in the ground state at $h=0$, and thus we obtain
after averaging with respect to the length $l$ with (\ref{distril})
the following a priori distribution
\begin{eqnarray}
P_{\Gamma}^{a priori} (m) \opsimeq \int_0^{+\infty} dl \frac{1}{l_{\Gamma} } e^{- \frac{l}{l_{\Gamma} }}
\frac{ e^{- \frac{m^2}{2 l} } }{\sqrt{2 \pi l}}
= \frac{1}{2 m_{\Gamma} } e^{- \frac{\vert m \vert }{m_{\Gamma} }}
\label{pmapriori}
  \end{eqnarray}
with the characteristic magnetization
\begin{eqnarray}
m_{\Gamma} = \sqrt{\frac{l_{\Gamma}}{2} } \opsimeq_{\Gamma \to 0}
\frac{1}{2 \Gamma^{\frac{\mu}{2} }} \left(  \frac{\mu}{  A  } \right)^{\frac{1}{2}} \label{mg}
 \end{eqnarray}

Now from this a priori distribution that describes 
the magnetizations of these domains in the $h=0$ ground state,
we wish to compute
the distribution of domain magnetizations obtained via
the renormalization procedure, where we have tried to flip intervals
in a iterative way.

A domain existing at scale $\Gamma$ was created at some previous scale
$\Gamma'$ representing the biggest of the couplings at the two boundaries.
Since the distribution of the already decimated coupling reads
\begin{eqnarray}
\rho_{\Gamma}^{small} (\vert J \vert)= 
\frac{ \theta(\vert J \vert <\Gamma ) \rho(\vert J \vert)}
{ \int dJ \theta(\vert J \vert <\Gamma ) \rho(\vert J \vert) }
 = \theta(\vert J \vert <\Gamma ) \frac{ \mu \vert J \vert^{\mu-1} }{ \Gamma^{\mu} }
 \end{eqnarray}
the distribution of the creation scale $\Gamma'$ is simply
\begin{eqnarray}
\rho_{\Gamma}^{creation} (\Gamma')= 
2 \rho_{\Gamma}^{small} (\Gamma') \int_{\vert J' \vert<\Gamma'}
d \vert J' \vert \rho_{\Gamma}^{small} (\vert J' \vert) =  \theta(\Gamma' <\Gamma ) \frac{ 2\mu (\Gamma')^{2\mu-1} }{ \Gamma^{2\mu} }
 \end{eqnarray}

The stability condition $m>-\frac{ \Gamma'}{h}$ at the scale $\Gamma'$
of its creation immediately yields the simple properties
for the probability distribution $P_{\Gamma}(m)$ of the domain magnetization
at scale $\Gamma$
\begin{eqnarray}
&& \theta \left( m >  \frac{\Gamma}{h} \right) P_{\Gamma}(m)
= 2 P_{\Gamma}^{a priori} (m)
\\
&& \theta \left(   -\frac{\Gamma}{h}>m \right) P_{\Gamma}(m)
=  0
\end{eqnarray}

For the values $\vert m \vert < \Gamma/h$, 
the probability distribution $\left[ P_{\Gamma}(m) \right]^{stable}$ induced by the only condition
to have been stable at the creation scale $\Gamma'$
reads
\begin{eqnarray}
\theta \left(   \frac{\Gamma}{h}>m>0 \right) \left[ P_{\Gamma}(m) \right]^{stable}
&& = P_{\Gamma}^{a priori} (m) 
\int_0^{\Gamma} d \Gamma' \frac{ 2\mu (\Gamma')^{2\mu-1} }{ \Gamma^{2\mu} }
\left[ 1+ \theta \left(m> \frac{\Gamma'}{h}\right)   \right] \\
&& = P_{\Gamma}^{a priori} (m) 
\left[ 1+  \left( \frac{ m h }{\Gamma}\right)^{2 \mu}   \right]
\label{stab1}
\end{eqnarray}
\begin{eqnarray}
\theta \left(   0>m>-\frac{\Gamma}{h} \right) \left[ P_{\Gamma}(m)
\right]^{stable}
&& =  P_{\Gamma}^{a priori} (m)
\int_0^{\Gamma} d \Gamma' \frac{ 2\mu (\Gamma')^{2\mu-1} }{ \Gamma^{2\mu} }
\left[ \theta \left(0>m>- \frac{\Gamma'}{h}\right) 
  \right]  \\
&& = P_{\Gamma}^{a priori} (m) 
\left[ 1 -  \left( \frac{ \vert m \vert h }{\Gamma}\right)^{2 \mu}   \right]
\label{stab2}
\end{eqnarray}

In the limit $\mu \to 0$, we have thus the simplification
that negative $m$ become negligible,
because two bonds weaker than $\Gamma$ are typically 
much weaker than $\Gamma$, as a consequence of the broadness of distribution.
So at leading order in $\mu$, we have the simple property

\begin{eqnarray}
\left[ P_{\Gamma}(m) \right]^{stable} \opsimeq_{\mu \to 0}
&& = \theta(m \geq 0) 2 P_{\Gamma}^{a priori} (m)  \\
&& = \theta(m \geq 0) \frac{1}{ m_{\Gamma} } e^{- \frac{\vert m \vert }{m_{\Gamma} }}
\end{eqnarray}

\subsection{ Probability measure for the fragmentation process at scale $\Gamma$}

The probability measure to find a bond of strength $\Gamma$ 
inside an interval $(L,M)$ existing at scale $\Gamma^-$
that becomes fragmented into two sub-intervals $(l_1,m_1)$
and $(l_2,m_2)$ reads
\begin{eqnarray}
&& {\cal N}_{\Gamma} (L,M ; l_1,m_1 ; l_2,m_2) d \Gamma dL dl_1 dl_2 dm_1 dm_2
\\ && = d \Gamma \rho_{\Gamma}(\Gamma) \frac{dL}{l_{\Gamma}} e^{- \frac{L}{l_{\Gamma}}} 
 dl_1 dl_2 \delta(L-(l_1+l_2))
 dM 2 \theta(M \geq 0)  dm_1 \frac{ e^{- \frac{m_1^2}{2 l_1} } }{\sqrt{2 \pi l_1}}
 dm_2 \frac{ e^{- \frac{m_2^2}{2 l_2} } }{\sqrt{2 \pi l_2}} \delta(M-(m_1+m_2))
\end{eqnarray}
Indeed, we have the following properties, for the integration over some variables.
After the integration over $(m_1,m_2)$, the distribution of $M$ is a Gaussian
as it should
\begin{eqnarray}
&& \int dm_1 \int dm_2 {\cal N}_{\Gamma} (L,M ; l_1,m_1 ; l_2,m_2)  = \\
&&  \rho_{\Gamma}(\Gamma) \frac{1}{l_{\Gamma}} e^{- \frac{L}{l_{\Gamma}}} 
  \delta(L-(l_1+l_2))
  2 \theta(M \geq 0)   \frac{ e^{- \frac{M^2}{2 L} } }{\sqrt{2 \pi L}} 
\end{eqnarray}
After the integration over all magnetizations $(m_1,m_2.M)$,
the distribution for $(l_1,l_2) $ is uniform except for the constraint
$l_1+l_2=L$
\begin{eqnarray}
&& \int dM \int dm_1 \int dm_2 {\cal N}_{\Gamma} (L,M ; l_1,m_1 ; l_2,m_2)  = \\
&&  \rho_{\Gamma}(\Gamma) \frac{1}{l_{\Gamma}} e^{- \frac{L}{l_{\Gamma}}} 
  \delta(L-(l_1+l_2))
\end{eqnarray}
After the integration over $(l_1,l_2)$, the probability to find a bond $\Gamma$ in an interval of length $L$
\begin{eqnarray}
 \int dl_1 \int dl_2 \int dM \int dm_1 \int dm_2 {\cal N}_{\Gamma} (L,M ; l_1,m_1 ; l_2,m_2)  = 
  \rho_{\Gamma}(\Gamma) \frac{1}{l_{\Gamma}} e^{- \frac{L}{l_{\Gamma}}} L 
\end{eqnarray}
is proportional to $L P_{\Gamma}(L)$ since they are $L$ possible positions.

In the following, when computing observables concerning the flips
at scale $\Gamma$, it will be more convenient to integrate first over the lengths
that play no direct role in the flip condition,
to keep the magnetizations that enters the flip condition
\begin{eqnarray}
{\cal N}_{\Gamma} (M ; m_1 ,m_2) \equiv && \int dL \int dl_1 \int dl_2 {\cal N}_{\Gamma} (L,M ; l_1,m_1 ; l_2,m_2) 
\\ && =  \rho_{\Gamma}(\Gamma) \frac{1}{l_{\Gamma}} 
  2 \theta(M \geq 0) \delta(M-(m_1+m_2))
\int_0^{+\infty} dl_1 e^{- \frac{l_1  }{l_{\Gamma}}} 
 \frac{ e^{- \frac{m_1^2}{2 l_1} } }{\sqrt{2 \pi l_1}}
\int_0^{+\infty} dl_2 e^{- \frac{l_2  }{l_{\Gamma}}} 
  \frac{ e^{- \frac{m_2^2}{2 l_2} } }{\sqrt{2 \pi l_2}}  \\
 && =  \rho_{\Gamma}(\Gamma)  
   \theta(M \geq 0) \delta(M-(m_1+m_2))
 e^{- \frac{\vert m_1 \vert }{m_{\Gamma} }}  e^{- \frac{\vert m_2 \vert }{m_{\Gamma} }}
\label{measure}
\end{eqnarray}

\subsection{ Flipping probability at scale $\Gamma$}

The number of bonds of strength $\Gamma$ that becomes frustrated at scale $\Gamma$
is proportional to
\begin{eqnarray}
N_{\Gamma}^{frus}(\Gamma) && 
= \int dM \int dm_1 \int dm_2 {\cal N}_{\Gamma} (M ; m_1 ,m_2)
 \left[ \theta \left(m_1<-\frac{\Gamma}{h}\right) 
+ \theta \left(m_2<-\frac{\Gamma}{h}\right)  \right] \\
&& = \rho_{\Gamma}(\Gamma)  
 \int dM  \theta(M \geq 0)  \int dm_1
 e^{- \frac{\vert m_1 \vert }{m_{\Gamma} }}  e^{- \frac{\vert M-m_1 \vert }{m_{\Gamma} }} 2 \theta \left(m_1<-\frac{\Gamma}{h}\right)  \\
&& = 2 \rho_{\Gamma}(\Gamma)  
 \int_0^{+\infty} dM  e^{- \frac{ M  }{m_{\Gamma} }}  \int_{\frac{\Gamma}{h}}^{+\infty} dm_1' e^{- 2 \frac{ m_1' }{m_{\Gamma} }}    \\
&& =   \rho_{\Gamma}(\Gamma) m_{\Gamma}^2 
 e^{- \frac{ \Gamma }{ h m_{\Gamma} } }
\end{eqnarray}
that should be compared with the total number of bonds of strength $\Gamma$ 
proportional to the normalization
\begin{eqnarray}
 N_{\Gamma}^{tot}(\Gamma) && = 
 \int dM \int dm_1 \int dm_2 {\cal N}_{\Gamma} (M ; m_1 ,m_2) \\
&& = \rho_{\Gamma}(\Gamma)  
 \int_0^{+\infty} dM   
\left[  \int_{-\infty}^0 dm_1
 e^{ \frac{ m_1  }{m_{\Gamma} }}  e^{- \frac{ M-m_1  }{m_{\Gamma} }}
+  \int_{0}^M dm_1
 e^{- \frac{ m_1  }{m_{\Gamma} }}  e^{- \frac{ M-m_1  }{m_{\Gamma} }}
+  \int_{M}^{+\infty} dm_1
 e^{- \frac{ m_1  }{m_{\Gamma} }}  e^{- \frac{m_1- M  }{m_{\Gamma} }}   \right] \\
&& = 2 \rho_{\Gamma}(\Gamma) m_{\Gamma}^2 
\end{eqnarray}

The flipping probability $F_{\Gamma}(\Gamma)$ of a bond $\Gamma$
at scale $\Gamma$ is given by the ratio of the two
\begin{eqnarray}
F_{\Gamma}(\Gamma)= 
\frac{ N_{\Gamma}^{frus}(\Gamma)  }{ N_{\Gamma}^{tot}(\Gamma)} =
 \frac{1}{2} e^{- \frac{ \Gamma }{ h m_{\Gamma}} } =  \frac{1}{2} e^{- x } 
\end{eqnarray}
where the rescaled variable 
\begin{eqnarray}
x = \frac{\Gamma}{h m_{\Gamma}} = \frac{2}{h} \left( \frac{A }
{ \mu }\right)^{1/2} \Gamma^{1 +\frac{\mu}{2} } 
\label{defx}
 \end{eqnarray}
represents the ratio between the quantity $\Gamma/h$ appearing in the flip condition
and the typical scale
$m_{\Gamma}$ of the magnetization of a domain existing at scale $\Gamma$.
At the beginning of the procedure $x \ll 1$, there is a finite probability of flip, of order $1/2$,
whereas for $x \gg 1$, the probability of flip becomes exponentially small.

\subsection{ Magnetization per spin at the end of the procedure}

To compute the magnetization per spin $m_{spin}(h)$,
we have to integrate over all the flips done at various scales,
and to keep track of the associated magnetization gain
\begin{eqnarray}
m_{spin}(h)  && = \sum_{\Gamma} \frac{ \sum_i \Delta m_i }{ \sum_i l_i }  =\int_0^{+\infty}  \frac{ (\Delta m)_{\Gamma,\Gamma+d\Gamma}}
{l_{\Gamma}}
\label{defms} 
 \end{eqnarray}
where $(\Delta m)_{\Gamma,\Gamma+d\Gamma}$ is the mean magnetization gain
associated to a domain flip at scale $\Gamma$, which can be expressed in terms of the measure (\ref{measure})

\begin{eqnarray}
&& (\Delta m)_{\Gamma,\Gamma+d\Gamma}
=  \int dM \int dm_1 \int dm_2 {\cal N}_{\Gamma} (M ; m_1 ,m_2) 
\left[ \theta \left(m_1<-\frac{\Gamma}{h}\right) 2 \vert m_1 \vert
+ \theta \left(m_2<-\frac{\Gamma}{h}\right) 2 \vert m_2 \vert \right] \\
&& = \rho_{\Gamma}(\Gamma)  
 \int dM  \theta(M \geq 0)  \int dm_1
 e^{- \frac{\vert m_1 \vert }{m_{\Gamma} }}  e^{- \frac{\vert M-m_1 \vert }{m_{\Gamma} }} 
\theta \left(m_1<-\frac{\Gamma}{h}\right) 4 \vert m_1 \vert  \\
&& = 4 \rho_{\Gamma}(\Gamma)  
 \int_0^{+\infty} dM  e^{- \frac{ M  }{m_{\Gamma} }}  \int_{\frac{\Gamma}{h}}^{+\infty}  dm_1' \  m_1'
 e^{- 2 \frac{ m_1' }{m_{\Gamma} }}  \\
&& =  \rho_{\Gamma}(\Gamma)  
 m_{\Gamma}^3  (1+2 x) e^{-2 x}
 \end{eqnarray}
where $x$ is the scaling variable defined in (\ref{defx}).

Finally, using $\rho_{\Gamma}(\Gamma)=2 A \Gamma^{\mu-1}$
and the new variable $x$ as integration variable instead of $\Gamma$, 
we obtain the 
magnetization per spin (\ref{defms}) as
\begin{eqnarray}
m_{spin}(h) = (1/2) \int_0^{+\infty}  
 d\Gamma \rho_{\Gamma}(\Gamma)   m_{\Gamma} 
\left[ 1+ 2 x \right]  e^{- 2 x }
 \end{eqnarray}
and the result
\begin{eqnarray}
m_{spin}(h) = h^{\frac{\mu}{\mu+2}}  \left( \frac{ 4 A }{\mu}  \right)^{\frac{1}{\mu+2}} c_{simple}(\mu)
\label{resmspin}
 \end{eqnarray}
The exponents in $h$ and $A$ agree with the exact results of \cite{gardner},
whereas
the prefactor reads
\begin{eqnarray}
c_{simple}(\mu) && = \frac{\mu }{ 2(\mu+2)} 
 \int_0^{+\infty}    dx x^{\frac{\mu}{\mu+2} -1} 
( 1+ 2 x )   e^{- 2 x }
=  = \frac{1+\mu}{2+\mu} 2^{- \frac{\mu}{\mu+2}} \Gamma \left( 1+ \frac{\mu}{\mu+2}  \right)
\\
&& = \frac{1}{2} +\mu \frac{1-\gamma_{Euler} -\ln 2}{4}
+ \mu^2 \left[ \frac{ (\gamma_{Euler} +\ln 2 )^2 }{16}
- \frac{12-\pi^2}{96} \right] + O(\mu^3)
 \end{eqnarray}
instead of the exact prefactor obtained via transfer matrix computations
\cite{gardner} that reads in our notations
\begin{eqnarray}
c_{exact}(\mu) = (\mu+2)^{-\frac{2}{\mu+2}}
(\mu+1) \frac{ \Gamma \left( \frac{\mu+1}{\mu+2} \right) }{\Gamma \left( \frac{1}{\mu+2} \right)}
= \frac{1}{2} +\mu \frac{1-\gamma_{Euler} -\ln 2}{4} +
\mu^2 \frac{ (\gamma_{Euler} +\ln 2 )^2 }{16}+ O(\mu^3)
\label{cexactmu}
 \end{eqnarray}
so the discrepancy with the exact prefactor (\ref{cexactmu})
only appears at order $\mu^2$.


\begin{thebibliography}{99}

\bibitem{imryma}
Y. Imry and S. K. Ma,
Phys. Rev. Lett. {\bf 35}, 1399 (1975).

\bibitem{us_rfim} 
 D. S. Fisher, P. Le Doussal and C. Monthus,  Phys. Rev. {\bf E64},
66107 (2001).

\bibitem{twolevel}
C. Monthus and P. Le Doussal, cond-mat/0407289.

\bibitem{chenma}
H.H. Chen and S.K. Ma, J. Stat. Phys. {\bf 29}, 717 (1982).

\bibitem{gardner}
E. Gardner and B. Derrida,  J. Stat. Phys. {\bf 39}, 367 (1985).

\bibitem{anderson72}
P.W. Anderson, B.I. Halperin, C.M. Varma, Phil. Mag. {\bf 25 }, 1 
(1972).

\bibitem{jerome}
J. Houdayer, private communication.

\bibitem{fragmentation}
 S.N. Majumdar and P.L. Krapivsky, Phys. Rev. {\bf E65}, 036127 (2002);
D.S. Dean and S.N. Majumdar, J. Phys. {\bf A35}, L501 (2002).

\bibitem{binderrfim}
I. Morgenstern, K. Binder and R.M. Hornreich,
Phys. Rev. {\bf B23}, 287 (1981).

\bibitem{binderinterface}
K. Binder, Z. Phys. {\bf B50}, 343 (1983); P. Shukla, cond-mat/0401253

\bibitem{rammal}
F. Barahona, R. Maynard, R. Rammal and J.P. Uhry,
J. Phys. {\bf A15}, 673 (1982). 

\bibitem{derrida}
B. Derrida and G. Weisbuch, Europhys. Lett. {\bf 4}, 657 (1987).

\bibitem{arcangelis} 
I.A. Campbell and L. de Arcangelis, Europhys. Lett. {\bf 13}, 587 (1990).

\bibitem{kobe}
T. Klotz and S. Kobe, J Phys {\bf A27}, L95 (1994).

\bibitem{domany}
G. Hed, A. K. Hartmann, D. Stauffer, and E. Domany
       Phys. Rev. Lett. {\bf 86}, 3148 (2001);
E. Domany, G. Hed, M. Palassini, and A. P. Young
       Phys. Rev. {\bf B64}, 224406 (2001);
G. Hed, A. P. Young, and E. Domany
       Phys. Rev. Lett. {\bf 92}, 157201 (2004).

\bibitem{Do_He}
E Domany, G. Hed, private communication.


\end{thebibliography}
\end{document}